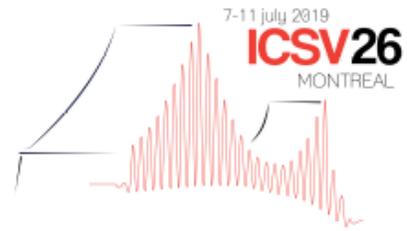

# SEAT PAN ANGLE OPTIMIZATION FOR VEHICLE RIDE COMFORT USING FINITE ELEMENT MODEL OF HUMAN SPINE

Raj Desai, Ankit Vekaria and Anirban Guha
*Department of Mechanical Engineering, IIT Bombay, India.*
*email: rajdesai@iitb.ac.in*

P. Seshu
*Department of Mechanical Engineering, IIT Dharwad, India.*

Ride comfort of the driver/occupant of a vehicle has been usually analyzed by multibody biodynamic models of human beings. Accurate modeling of critical segments of the human body, e.g. the spine requires these models to have a very high number of segments. The resultant increase in degrees of freedom makes these models difficult to analyze and not able to provide certain details such as seat pressure distribution, the effect of cushion shapes, material, etc. This work presents a finite element based model of a human being seated in a vehicle in which the spine has been modelled in 3-D. It consists of cervical to coccyx vertebrae, ligaments, and discs and has been validated against modal frequencies reported in the literature. It was then subjected to sinusoidal vertical RMS acceleration of 0.1 g for mimicking road induced vibration. The dynamic characteristics of the human body were studied in terms of the seat to head transmissibility and intervertebral disc pressure. The effect of the seat pan angle on these parameters was studied and it was established that the optimum angle should lie between 15 and 19 degrees. This work is expected to be followed up by more simulations of this nature to study other human body comfort and seat design related parameters leading to optimized seat designs for various ride conditions.

## 1. Introduction

One of the most critical selection criteria while designing a vehicle is ride comfort. It is affected by parameters such as road irregularities, whole body vibration, sitting posture, suspension system, etc. So, it is very important to analyze the human body's response to these random vibrations. The human body is a system whose biomechanical properties vary from time to time and from one part of the body to another. A driver is exposed to low-frequency vibrations while driving which are produced due to the interaction between vehicle and road. Biomechanics has played a major role in biomedical science. There has been a significant contribution from both engineers and physicians when it comes to dealing with human body analysis. In this emerging area of bio-medical science, various parts of the human body are studied and analyzed through different methods. Whole body vibration is one area in which skeleton injuries and its joints including spine are studied. The human spine is a dynamic mechanical structure which supports the loads and bending moments of the head, upper torso and any weights that are lifted by pelvis [1]. Because the spine carries the load, there have been many instances of spinal injuries like back pain among vehicle drivers exposed to vibration. Thus, it is essential to understand how to decrease the intensity of such vibrations and its adverse effects on the human body. The spine is made up of 33 bones attached on top of the other. This spinal column provides the main support to the torso, allowing the body to stand straight, bend, twist and at the same time protects the spinal cord from injury. Various parts of the spine are affected by injuries, and it has become very important to understand the human spine as spine supports upper body movement. Of the various parts of the spinal column, cervical vertebrae is located at the posterior of the neck, and its function is to provide head support and movement. There are a total of seven cervical vertebrae named as $C_1$ to $C_7$. They allow that neck can move freely. The thoracic vertebrae are stacked together in the upper trunk of the vertebral spine. There are twelve thoracic vertebrae namely $T_1$ to $T_{12}$. They support ribs and the upper mass of the body. They protect the fragile spinal cord, and it runs through the vertebral canal [2]. The thoracic vertebrae are located in the thorax posterior and medial to the ribs. Lumbar vertebrae withstand the weight of the torso. There are five lumbar vertebrae numbered $L_1$ to $L_5$. These vertebrae are larger in size and they can bear the stress during the lifting heavy objects. Sacrum is a triangular bone located at the lower back between hip bones and pelvis. It connects the spine to the hip bones. There are five sacral vertebrae and they are fused together. The last portion is called coccyx in which four fused tail bones are attached to the ligaments and muscles of the pelvis. The vertebra in the spinal column are separated by intervertebral discs. They provide cushioning to the vertebrae preventing the bones from rubbing together. Intervertebral discs consist of nucleus pulposus and annulus fibrosus[3]. Nucleus pulposus is at the core of the disc. It is a jelly like material that allows the vertebral discs to carry forces of compression and torsion. The annulus fibrosus lies outside the intervertebral disc and covers the soft inner core (nucleus pulposus). The length of the spinal cord is approximately 18 inches and its thickness same as that of the thumb. It passes through the spinal canal. The cord fibers are separated at the end and continue towards the tailbone, and branching off to the legs and feet. Two vertebrae are connected by facet joint and they help the vertebrae in bending and twisting. Facet joints have cartilage which allows the vertebrae to move smoothly against each other. They have very strong fibrous bands which hold the vertebrae together, give stability to the spine and protect the discs. From this description, it is obvious that it is very difficult for a rigid body-based model to capture all the elements of the spine. Only a finite element-based modelling method can capture the spine in all its complexity and analyze the stress, strain and vibration response in every section. This work has used such a model.

The primary focus of this work is a vibration analysis of the spine model to improve human comfort on the vehicle. The response of the human spine is limited to vertical sinusoidal vibration. A CAD model of the spine is analyzed by the finite element method in ANSYS, and the results are verified with the experimental data taken from the literature. The resonant frequencies of the spinal model and its effects are studied. The vertical transmissibility of the spine is studied and analyzed for sinusoidal vertical vibration. After that, seat-to-head transmissibility (STHT) and intervertebral disc pressure have been studied to obtain an optimum seat-pan angle for maximum comfort.

## 2. Methods of human body Modelling

There are a large number of biodynamic models for the study of biodynamic responses of a seated human being to vibration. These models can be divided into three main categories - lumped parameter, finite element and multibody models [4].

## 2.1 Lumped parameter system

In lumped parameter models, the human body is considered as several concentrated rigid point masses which are inter connected by springs with dampers. These masses represent body parts. So, the whole system is made up of masses, springs, and dampers as shown in Fig. 1 and parameter values in Table 1. This is the most commonly used analytical method and is very easy for validation. Its primary limitation is that it has been used only for one-dimensional vibration analysis.

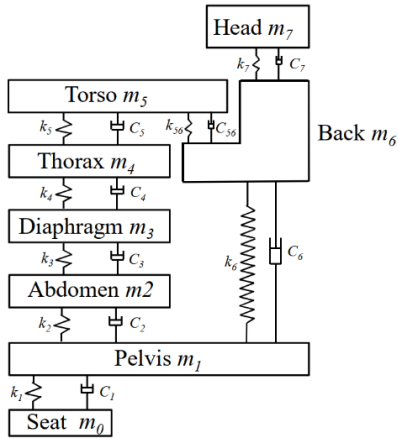

Figure 1. Seven degree of freedom lumped parameter model [5]

Table 1: Model Parameters

| Mass (kg) | Stiffness(N/m) | Damping(N-s/m) |
|---|---|---|
| $m_1$=27.230 | $k_1$=25000 | $c_1$=371 |
| $m_2$=5.921 | $k_2$=877 | $c_2$=292 |
| $m_3$=0.455 | $k_3$=877 | $c_3$=292 |
| $m_4$=1.356 | $k_4$=877 | $c_4$=292 |
| $m_5$=32.762 | $k_5$=877 | $c_5$=292 |
| $m_6$=6.820 | $k_6$=52600 | $c_6$=3580 |
| $m_7$=5.450 | $k_7$=52600 | $c_7$=3580 |
| | $k_{56}$=52600 | $c_{56}$=3580 |

## 2.2 Multibody model

In multibody (MB) models, the human body is considered as rigid bodies usually interconnected by revolute pairs (though other lower pairs have also been used). In a few instances, the rigid bodies are connected by bushing elements. All the elements have either rotational or translational motion or both. These models can be further classified as kinetic and kinematic models. The kinetic models include the motion of each and every segment of body and forces acting on them while the kinematic models ignore the forces. Amirouche and Ider [6] created an MB model that has 13 rigidly connected links and flexible segments internally connected by revolute, spherical and free joints. Cho and Yoon [7] developed a 2D model to analyze vehicle ride comfort. Pennestri [8] created a numerical model based on the dynamics of MB. They conducted tests on a vehicle to validate static and dynamic forces. This approach allows the study of the behaviour of the body in 3D. The simulation was done with different poses such as seating and standing positions with a change in the various parameters like inclination of seat and height of the seat. Liang and Chiang[9] mentioned two models in a multibody system to study the biodynamic behavior for various postures. Their main intention of the study was to study the response of the human body by considering the effect of back support. Kim et al. [10] developed a multibody model with pin joints which are non-linear having multi DoF as shown in Fig.2 and model properties in

Table 2.

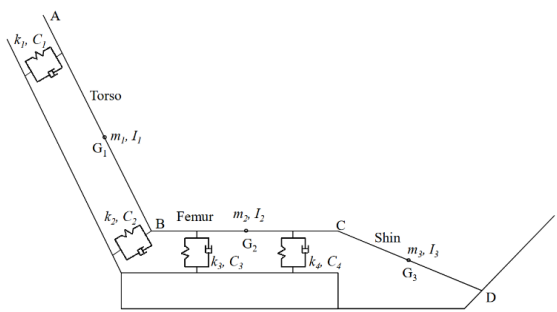

Figure 2. Kim multibody model [10].

Table 2: Model Parameters

| Mass (kg) | Stiffness(N/m) | Damping(N-s/m) |
|---|---|---|
| $m_1$=3.988 | $k_1$=61106.4 | $c_1$=17.62 |
| $m_2$=34.13 | $k_2$=276708.0 | $c_2$=0.0 |
| $m_3$=17.84 | $k_3$=126140.6 | $c_3$=598.98 |
| $m_4$=7.948 | $k_4$=39470.7 | $c_4$=48.95 |

Though MB models are an improvement over lumped parameter models, even they cannot capture all the complexities of a human body. As a result, some important details such as pressure distribution and the effect of different cushion shapes cannot be studied.

### 2.3 Finite Element Method modelling

In finite element models, the human body consists of innumerable elements whose mechanical properties are established from various experiments on a human cadaver. Each of these elements can be modeled with a fairly high level of detail to match the corresponding parts of the human body. These can then be used to study the biodynamic vibrational response of a human body and can be used in predicting injuries to passengers during an accident or normal driving to a level of detail which is not possible in lumped parameter or MB models [11]. To take a specific example, intervertebral disc pressure, a key marker for back pain, can be accurately modelled only in finite element models. Many experimental studies have been reported on the response of the human body subjected to vibration to obtain the resonant frequency of different regions of the human body. Such identification of resonant frequencies helps in deciding human comfort level when the body is subjected to vibration. In the vertical vibration direction, for both sitting and standing postures, the resonant frequency was measured in the range of 4–8 Hz [12]. It was also noted that seated body with vibration at 5 Hz increases spinal injury [13]. Kasra [14] found a resonant frequency of 26 Hz by both experimentally and analytically using one motion segment of the lumbar spine. Goel [15] identified a resonant frequency of 17.8 Hz using two segmental motion model by finite element analysis. However, the in vivo resonant frequency is in the range of 4 to 8 Hz [16]. Hence, the use of segmental models is not recommended. This again establishes the need for a finite element analysis of the complete spine and validate the resonant frequency and transmissibility results with experimental data. Such a model can be used for predicting the intervertebral disc pressure and study how it changes with the seat pan angle. This can lead to a suggestion for the optimum angle. To the best of our knowledge, such a study has not been reported in the literature.

## 3. Properties of the human biodynamic model

The human body is a compound mechanical system. It responds differently for different frequencies of applied vibrations as each region has its own resonance behavior. At the low-frequency vertical vibration of 1 Hz, the response of seat and the body parts are very similar as there is very small relative motion. The relative motion of the body increases with increasing frequencies and reaches a maximum at the resonance frequency. Thus, resonance occurs when transmissibility reaches its peak. The high amplitude of vibration at resonance can adversely affect different physiological functions like organ function, muscle-ligament function, respiratory function and blood circulatory function.

Table 3: Material properties of components of the spine

| Parts | Young's Modulus (MPa) | Poisson ratio |
| --- | --- | --- |
| **Vertebrae** | | |
| Cortical Bone | 12000 | 0.3 |
| Cancellous bone | 100 | 0.2 |
| Endplate | 12000 | 0.3 |
| Posterior bone | 3500 | 0.25 |
| **Intervertebral disc** | | |
| Annulus ground substance | 4.2 | 0.45 |
| Nucleus pulposus | 1 | 0.4999 |

Different body parts have different resonance frequencies. The spinal column is made up different elements having distinct material properties. Material properties of those elements are listed in Table 3 and dimensions, and mass densities of different parts of the spine are listed in Table 2 [17]. These properties have been used in the modelling of the spine.

Table 4: Dimensions and mass densities of the components of the spine

| Parts | Section type | Shell Thickness(mm) | Density (g/cm$^3$) |
|---|---|---|---|
| Annulus ground sub-stance | Solid | | 1.05 |
| Annulus fiber laminate | Membrane | 1.5 | 1 |
| Cancellous bone | Solid | | 1.1 |
| Cortical bone | Membrane | 0.35 | 1.7 |
| Endplate | Membrane | 0.15 | 1.7 |
| Nucleus pulposus | Solid | | 1.02 |
| Posterior bone | Solid | | 1.4 |
| Ligament | Truss | | 1 |

## 4. Analysis of spine

A three-dimensional finite element model of the entire human spine was created. It was subjected to sinusoidal vertical RMS acceleration of 0.1 g. Out-of-plane (perpendicular to sagittal plane) displacements (translational or rotational motion) was not allowed for any node. A ligamentous model was created, and the weight of the spine and the muscles associated with it were taken into consideration. The model contains 33 elements of vertebrae as shown in Fig. 3a.

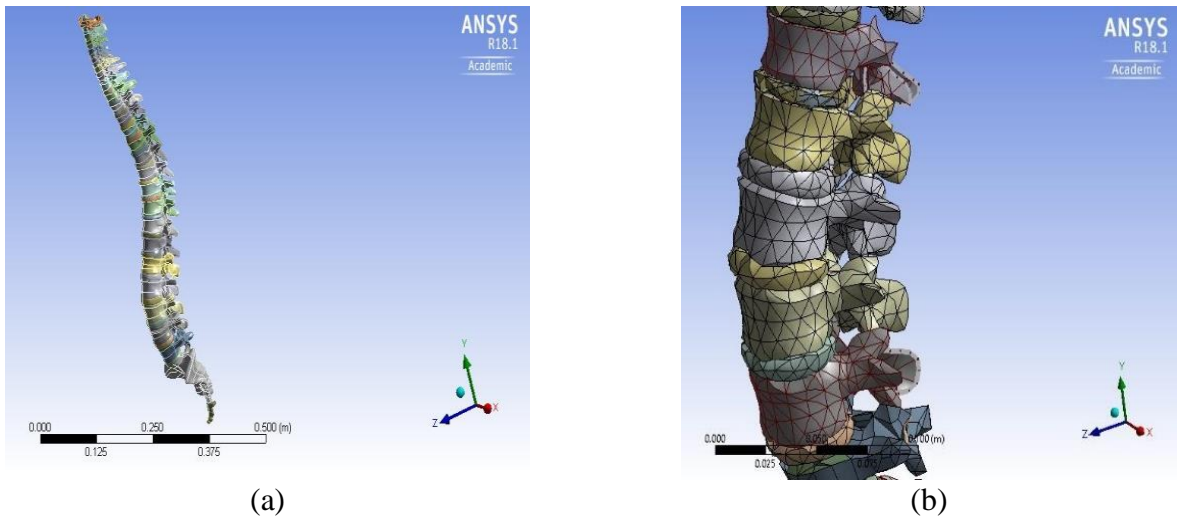

(a) (b)

Figure 3. A three-dimensional model of the spinal column

Boundary and loading conditions were such that sacrum of the spinal segment was fixed in all directions. A sinusoidal vertical load varying between 360 N and 440 N was applied at the top to simulate sinusoidal loading and mass of the head and upper torso. The FEM software package Ansys was used. Meshing was created in the spine as shown in Fig. 3b. Meshing model has different elements types within the finite element model such as 3D tetrahedral and pyramid element each of 4 nodes and 5 nodes respectively. The resonant frequency for the first four modes were studied. The first resonant frequency was 8.9 Hz. The other frequencies and the corresponding values obtained from literature in [18] are mentioned in Table 5 and show a close match between the two. The very first mode of vibration is in the anteroposterior direction. The other end of the head region goes away from the longitudinal axis. In the second mode, the spine moved in both the anteroposterior as well as vertical directions.

Table 5: Value of First Four Natural Frequencies.

| Motion Segment | First Four Natural Frequencies (Hz) -Model | Literature [18] |
|---|---|---|
| Head to sacrum model | 8.9 | 8.32 |
|  | 6.12 | 5.34 |
|  | 2.6 | 2.79 |
|  | 0.35 | 0.43 |

## 5. Vehicle seat-pan angle

The seat-pan angle of the vehicle is one of the most significant criteria while designing any vehicle seat because it directly affects the comfort and posture of the driver. A vehicle traveling on roads vibrates due to the road irregularities. These vibrations are transferred to the chassis through the tires. From the chassis, these vibrations are transferred to the vehicle body parts, including the seat. As the human body is in direct contact with the seat cushion, it is very important that the seat should provide good vibration isolation so that vibrations of the human body can be minimized. Sitting posture is the most easily controllable factor for attempting vibration isolation, and that is dependent primarily on a seat-pan angle. The vibratory excitation of the human body occurs primarily through the seat and depends on the seat-pan angle. Thus, the seat-pan angle is a very important parameter for analysis of ride comfort.

### 5.1 Effect of seat-pan angle on seat-to-head transmissibility

The STHT is the ratio of the biodynamic response motion of the human head to the forced vibration motion at the seat-pelvis interface as shown in Eq. (1).

$$STHT(f) = \frac{Head\ output\ acceleration\ (f)}{Seat\ input\ acceleration(f)} \quad (1)$$

STHT is a 'through the body' biodynamic function which signifies the extent to which the input vibration is transmitted to the parts of the body. The seat to head transmissibility as shown in Fig. 4 plays a vital role in deciding any biodynamic response of the body as one can find how head acceleration is amplified or damped as compared with input excitation over the entire frequency range (0 to 20 Hz). The trend of this plot is close to that of similar plots obtained from experiment [9] and this further validates the FEM model developed in this work. This section involves the effect of seat-pan angle on STHT. The angle of the seat was changed from zero to 40° for input acceleration of 0.1 g. The results obtained after the simulation using ANSYS for peak in STHT over 0 to 20 Hz frequency range across 0 to 40° seat pan angle has been summarized in Fig. 5.

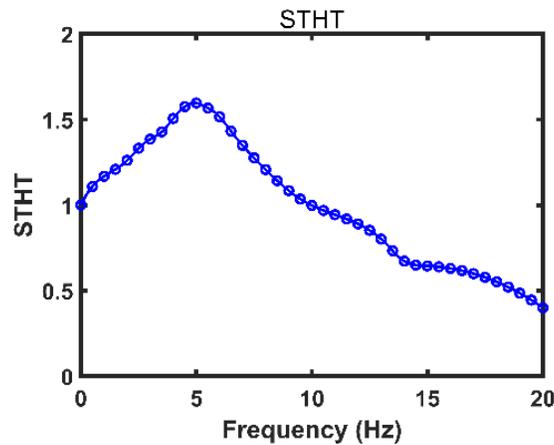

Figure 4. Seat to head transmissibility frequency response (seat pan angle 0°).

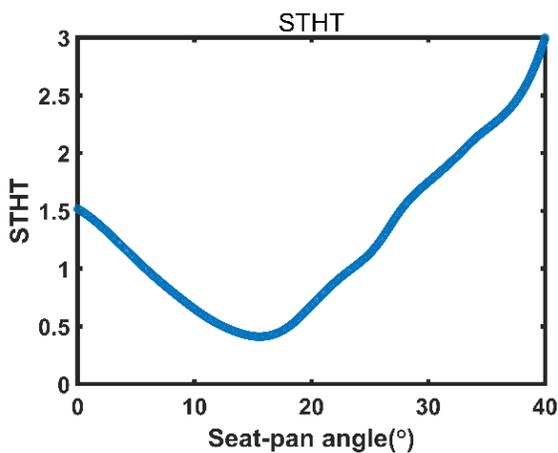

Figure 5. Response of STHT at different seat-pan angle.

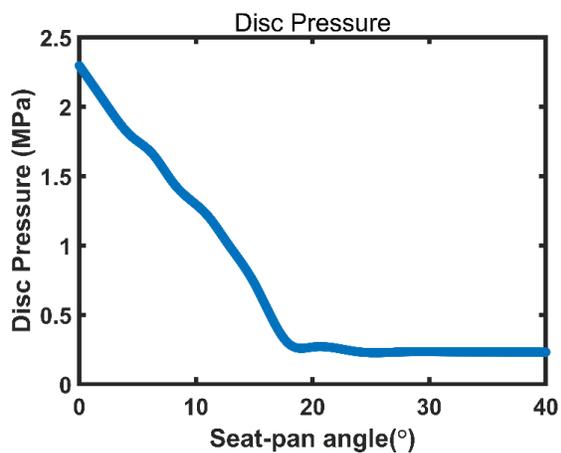

Figure 6. Response of disc pressure at different seat-pan angle

### 5.2 Intervertebral disc pressure at different seat-pan angles

The intervertebral disc pressure plays a vital role in deciding susceptibility of the body to pain. This chapter involves the effect of seat-pan angle on maximum disc pressure of the human spine. In this study, the angle of the seat was changed from zero to 40°. The results obtained after the simulation using ANSYS has been summarized in Fig. 6. The angle is measured from a position in which the seat is horizontal.

## 6. Results

As seen from Fig. 4, that the plot of STHT versus seat-pan angle is an inverted bell-curve. It drops to a minimum at 14.8° of seat-pan angle and then starts increasing. So, this is the seat-pan angle at which human being should be most comfortable from the point of view of STHT. A similar plot for intervertebral disc pressure is shown in Fig. 6. The values drop till 18.7° and then remain nearly constant. So, a consideration of both STHT and intervertebral disc pressure leads to the conclusion that the range of 15 to 19° should be studied further for determining the optimum seat pan angle. Other human body comfort related parameters may be studied in this range and that may lead to the truly optimum seat pan angle.

## 7. Conclusion

This paper presents a study in which an attempt has been made to identify the optimum seat pan angle for maximum comfort to a seated human being in a vehicle. Of the different modelling techniques reported in the literature for this purpose, the finite element method has been chosen for its ability to capture the intricacies of the human body - specifically the intervertebral disc pressure of the spine. This parameter was chosen due to its correlation with susceptibility of back pain. However, vibration isolation is expected to be better correlated to STHT. So, this too was studied. The finite element model was validated against the first four fundamental frequencies reported in the literature. It was then subjected to sinusoidal excitation mimicking the vibration due to road imperfections. Simulations were done by changing the seat pan angle from zero to 40°. It was found that the STHT is minimum at a seat pan angle of 14.8° and the intervertebral disc pressure reaches a minimum at 18.7° and after that remains constant till 40°. This leads to the conclusion that the ideal seat pan angle should be between 15 and 19 degrees and that a study of other human body comfort parameters in this range is necessary to obtain the true optimum. Other parameters such as cushioning effects and other postures of seating in a vehicle can also be included in a comprehensive study of this topic.